\pdfoutput=1

\documentclass[11pt]{article}

\usepackage[]{ACL2023}

\usepackage{times}
\usepackage{latexsym}

\usepackage[T1]{fontenc}

\usepackage[utf8]{inputenc}

\usepackage{microtype}

\usepackage{inconsolata}

\usepackage{url}
\usepackage{graphicx}
\usepackage{multirow}
\usepackage{booktabs}
\usepackage{hyperref}
\usepackage{enumitem}
\usepackage{tikz}
\usetikzlibrary{shapes.misc, calc}

\title{SAINE: Scientific Annotation and Inference Engine of Scientific Research}

\author {
    Susie Xi Rao\textsuperscript{\rm 1},
    Yilei Tu\textsuperscript{\rm 2},
    Peter H. Egger \textsuperscript{\rm 3} \\
    \textsuperscript{\rm 1} Chair of Applied Economics; Institute of Computing Platforms, ETH Zurich\\
    \textsuperscript{\rm 2} Chair of Applied Economics; Department of Computer Science, ETH Zurich\\
    \textsuperscript{\rm 3} Chair of Applied Economics, ETH Zurich; CEPR; CESifo; \\ Leverhulme Centre for Research on Globalisation and  Economic Policy (GEP), \\ University of Nottingham \\
    \{srao, yiletu, pegger\}@ethz.ch
}

\begin{document}
\maketitle
\begin{abstract}

We present \textbf{SAINE}, a \textbf{S}cientific \textbf{A}nnotation and \textbf{I}nference \textbf{EN}gine based on a set of standard open-source software, such as Label Studio and MLflow. We show that our annotation engine supports the development of classification systems. Based on our previous work on hierarchical discipline classifications, we demonstrate SAINE's use for understanding the space for scholarly publications. The user study of our annotation results shows how the user input processed with SAINE can support the classification process. SAINE can be particularly useful in supporting downstream meta-science research. We welcome collaboration and feedback from the scientific community on these projects. The demonstration video can be accessed from \url{https://youtu.be/yToO-G9YQK4}. 
A live demo website is available at \url{https://app.heartex.com/user/signup/?token=e2435a2f97449fa1} upon free registration. 
\end{abstract}

\section{Introduction}
\begin{figure*}
    \centering
    \includegraphics[scale=0.4]{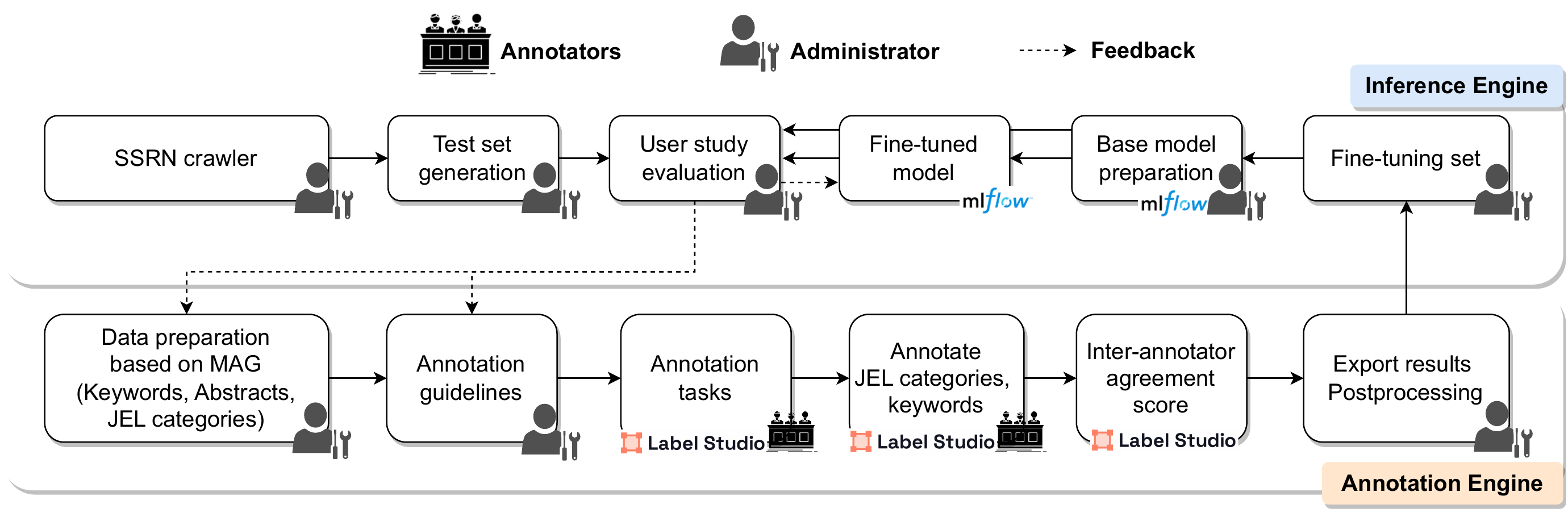}
    \caption{SAINE Workflow and Pipeline.}
    \label{fig:pipeline}
\end{figure*}
A precise classification of publications across and within disciplines is key not only for a fast and comprehensive search to guide researchers to relevant material but also to identify the novelty of research, the standing and significance of scholars, and of the relative growth of fields of work.

Machine learning develops into being not only \textit{a} but \textit{the} customary approach to establish such a classification. Clearly, one would expect a search that is geared towards identifying a high-quality corpus of keywords to benefit crucially from supervision. Existing classifications of academic output are based on a blend of (supervised) author-chosen and (unsupervised) machine-chosen keyword lists, where the composition of the blend is unknown to the researcher. 

Prevailing systems of keywords for academic publications are lists based on abstracts in a discipline, field, and subfield,  distilled from
\begin{itemize}[leftmargin=*, itemsep=0pt, parsep=0pt, topsep=0pt]
\item unsupervised machine learning (from word or phrase frequencies);
\item supervised learning (mostly from keyword self-reporting by authors);
\item semi-supervised learning (a mixture of the two; e.g., as done by Microsoft Academic Graph (MAG) described in~\citet{sinha2015overview, 10.3389/fdata.2019.00045, wang2020microsoft}).
\end{itemize}

For designing an annotation and inference engine that helps establishing a classification system of scientific publications, one would target developing a tool with the following features: (1) a simple user interface with clear annotation instructions; (2) a reproducible pipeline across various disciplines; (3) good support for inference tailored to downstream tasks (e.g., model retraining) in meta-science studies. 

Among the existing open-source annotation tools, Label Studio \cite{LabelStudio} suits those needs. Note that \citet{gayoso2019review} have reviewed extensively the annotation tools that facilitate classification tasks. However, the reviewed tools are either not open-sourced or are domain-specific and, hence, do not share the aforementioned targeted features. 

In this system demonstration, we utilize a set of standard open-source software, mainly Label Studio \cite{LabelStudio}, \href{https://mlflow.org/}{MLflow} and \href{https://fastapi.tiangolo.com/}{FastAPI} to configure an annotation and inference engine for scientific publication annotations. In this demonstration, we illustrate the benefit of using supervised learning based on pre-established keyword lists and abstracts, and how annotators can help us better understanding the importance of \textit{supervised} learning in establishing a classification of academic publications. 

This system is built on top of the hitherto largest-scale \textit{multi-class} hierarchical classification study across all academic research disciplines in both \textit{single-label} and \textit{multi-label} settings (cf.~\citet{rao2023hierarchical}). There, we have built a supervised hierarchical classification system that associates every publication with at least one and potentially several disciplines, fields, and subfields. 

With the annotations above, we conduct a small user study with domain experts using our \textbf{annotation} engine. We then invoke our \textbf{inference} engine to fine-tune the base models in \citet{rao2023hierarchical}. The comparison between the base and fine-tuned models shows that the proposed annotation and inference system is able to benefit the development of more accurate classifications. 

To summarize, the paper presents a scientific annotation and inference engine called SAINE, which is based on open-source software like Label Studio and MLflow. \textbf{The main contributions of the paper are:}
(1) The demonstration of using SAINE in understanding the space for scholarly publications, particularly in hierarchical discipline classifications.
(2) The result of a user study, which shows that user input collected with the help of SAINE can help better understand the classification process.
(3) The ability of SAINE to benefit the further development of a more accurate classification, demonstrated through the comparison between the base and fine-tuned models.
(4) The potential of SAINE to support downstream meta-science projects and foster greater transparency and understanding of scientific research.

Overall, the paper presents the benefits of supervised learning and the importance of having a simple user interface with clear annotation instructions, reproducible pipelines, and good support for inference in scientific publication annotations. The \href{https://app.heartex.com/user/signup/?token=e2435a2f97449fa1}{live demo website} and \href{https://youtu.be/yToO-G9YQK4}{demonstration video} are also available for those interested in further exploring SAINE.\footnote{Note that the annotation engine is used for internal testing at the moment but will open for beta testing soon. We have 75 users queuing for the beta stage. We are working on incorporating user profiles either through self-declaration or automatic screening.} The codebase for development is publicly available \href{https://gitlab.ethz.ch/raox/science-clf/-/tree/labelstudio}{under this link} and collocates with the codebase of \citet{rao2023hierarchical}.

In Figure~\ref{fig:pipeline} we illustrate the workflow in SAINE by assigning the roles of "Administrator", "Annotators", "Label Studio", and "MLflow" to each task in the pipeline. The sections are organized as follows. Section~\ref{sec:labelstudio} introduces the functionality of Label Studio and its fit to our annotation needs as well as our annotation guidelines for experts. Section~\ref{sec:implementation} specifies the annotation design for the field of \textit{Economics} and discusses the annotation results. Section~\ref{sec:inf} discusses the integration of annotation results into the pre-trained base models and fine-tuned ones with MLflow. We devote Section \ref{sec:annotation-efficiency} to discuss our preliminary experiments on improving annotation efficiency. We then conclude this system demonstration with a discussion of system limitations, ethics, and broader impact statements. 

\section{Annotating Scientific Articles with Label Studio}
\label{sec:labelstudio}
We briefly introduce functionalities of Label Studio and illustrate why Label Studio is a suitable tool to annotate scientific articles. \textbf{The goals of annotations in the present context are three:} 
\textbf{(1) [To judge the appropriateness of an assigned category.]} Experts judge whether a scientific article has a correctly assigned category. If a category does not suit the abstract's content, the annotator will pick a new category from a pre-defined list. 
\textbf{(2) [To evaluate keywords.]} Experts evaluate keywords assigned by Microsoft Academic Graph (MAG) and mark missing keywords in the abstract.
\textbf{(3) [To calculate inter-annotator agreement.]} The annotation engine should permit an efficient way of calculating inter-annotator agreement (IAA) scores among annotators.

\subsection{Label Studio}
Label Studio is a powerful and versatile annotation tool that can handle various types of annotation tasks. Here are some of the features of Label Studio that make it suitable for the annotation tasks.

\begin{itemize}[leftmargin=*, itemsep=0pt, parsep=0pt, topsep=0.5pt]
    \item \textbf{[Customizable interface.]} Label Studio allows to design a customized annotation interface. One can create a pre-defined list of categories for the experts to choose from and provide them with clear instructions on how to evaluate the assigned category.
    \item \textbf{[Multiple annotation types.]} Label Studio supports various types of annotation, including text classification, entity recognition, and sequence labeling. Therefore, different types of annotation can be used to evaluate keywords, mark missing keywords, and judge predicted categories.
    \item \textbf{[Collaboration and version control.]} Label Studio enables multiple experts to work on the same project simultaneously, allowing efficient and collaborative annotation. It also includes a version control system that tracks changes to the annotations, facilitating easy comparison and IAA evaluation.
    \item \textbf{[Inter-annotator agreement (IAA).]} Label Studio has built-in tools to calculate IAA scores. These tools can help evaluate the consistency and reliability of chosen annotations.
    \item \textbf{[Integrations with machine-learning models.]} Label Studio also provides integration with various machine learning (ML) models. Although we do not use the integrated ML functions, Label Studio allows us to export the annotation results in JSON, with which we improve the classification models using the annotated data in the inference engine.
\end{itemize}

Overall, Label Studio offers a powerful and customizable annotation platform that can handle relevant annotation tasks, facilitate efficient collaboration among experts, and efficiently compute IAA.

The project manager uses an administrative panel (Figure~\ref{fig:labelstudio_admin} in Appendix \ref{app:more-function}) to assign annotation tasks to each registered annotator and can monitor the annotation progress. The manager can also adjust the assigned annotations based on individual progress, as well as inspect tasks by annotation progress and IAA metrics.  

\subsection{Annotation Guidelines}
When a publication is annotated, each annotator is provided with the abstract, the keywords offered by MAG, and the assigned category based on the keywords provided by MAG. The categories of a discipline classification (such as the Journal of Economic Literature, JEL, clasification in economics) are assigned to MAG publications on the basis of the keywords. Therefore, MAG's keywords help us identify potential misalignments and better understand the classifiers we built.

\begin{figure*}[!t]
    \centering
    \includegraphics[width=0.8\linewidth]{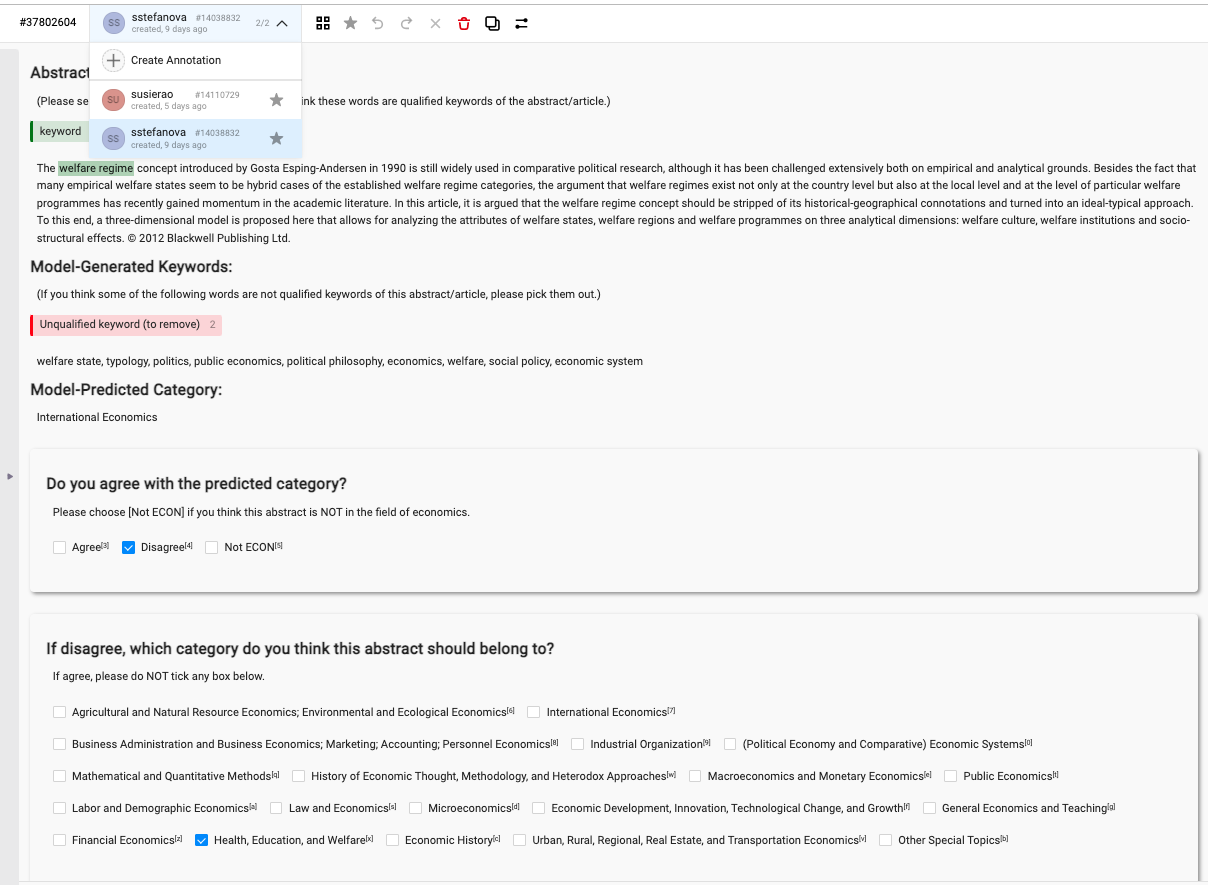}
    \caption{Publication Annotations by Multiple Annotators.}
    \label{fig:labelstudio_multipleannotator}
\end{figure*}

The annotation samples provided in the annotation engine are stratified sampled (ratio:~2e-5) across all classes of the training set introduced by \citet{rao2023hierarchical} for one discipline. Each annotator is required to judge whether a category is correctly assigned to an abstract. If not, the annotator is required to select the suitable one from a predefined list. The annotator is also required to evaluate MAG-generated keywords and make corrections (by removing unqualified keywords/marking suitable keywords from the abstract). Figure~\ref{fig:labelstudio_multipleannotator} shows two annotations of one publication. Label Studio makes it easy to navigate among the annotations generated by various annotators on an identical instance. Note that, as we discussed in \citet{rao2023hierarchical}, our multi-class hierarchical classification system is modularized in both \textit{single-label} and \textit{multi-label} settings.\footnote{We offer an brief overview of our hierarchical classification system in Appendix \ref{app:sys}.} The current annotation engine is equipped with both annotation functionalities. For the sake of system demonstration and user study in Section~\ref{sec:implementation}, we discuss the \textit{single-label} setting. More details on the multi-label setting are provided in Appendix~\ref{app:multi-label}.

\section{Implementation: User Study in \textit{Economics}}
\label{sec:implementation}

We now use \textit{Economics} as a discipline to show how we utilize the annotation engine to collect expert annotations.  

\subsection{Annotation Design}
We invited three economist experts from the Chair of Applied Economics at ETH Zurich to join the annotation project by accessing \href{https://app.heartex.com/user/signup/?token=e2435a2f97449fa1}{this link}. Annotation guidelines are given at \href{https://gitlab.ethz.ch/raox/science-clf/-/blob/labelstudio/labelstudio/annotation_guideline.md}{here}. Of the three experts, one has annotated all provided instances (Annotator 1), one has annotated 10\% of the instances (Annotator 2), and one has annotated a subset of instances with an \textit{ex ante} denomination in Urban and Spatial Economics only (Annotator 3). Each annotator received a user panel like Figure~\ref{fig:labelstudio_annotator} in Appendix~\ref{app:more-function}.

\subsection{Annotation Results in Label Studio}

Altogether, 788 instances of abstracts and keywords from MAG had to be annotated for a \textit{single-label} classification. In Economics, a standardized field and subfield system with keywords exists, and it is called \href{https://www.aeaweb.org/econlit/jelCodes.php?view=jel}{the Journal of Economic Literature (JEL) classification system}. This system is known to all academic economists and serves as a guiding principle to associate an article or a topic with a specific subfield in \textit{Economics}. The subfields in the JEL categories are associated with keywords.

We report the annotation time and IAA scores that are automatically calculated by Label Studio (see \href{https://labelstud.io/guide/stats.html}{the official documentation} for the steps). The final task agreement score is calculated by averaging all IAA scores for each annotation pair. Table~\ref{tab:iaa} illustrates the IAA scores amongst three experts. Annotators 1, 2, and 3 have annotated 788, 181, and 99 instances, respectively. The annotation overlap between the pairs of annotators is 4 or 7\% of the overlapping instances (Annotators 2 and 3), 99 or 100\% of the overlapping instances (Annotators 1 and 3), and 181 or 100\% of the overlapping instances (Annotators 1 and 2). The median annotation time of Annotators 1-3 per instance was 17.7s, 29.8s, and 40.9s, respectively. The annotators were entitled to disapprove of the assigned category based on MAG upon suggesting an alternative category. Marking and filling in missing keywords is time consuming, reading the MAG-generated keywords can help to some extent the annotation speed. However, all annotators reported that the MAG-provided keywords could be a source of error for wrongly assigned categories. As discussed among the annotators after they underwent the annotations separately, the category they found the best was for \textit{Mathematical \& Quantitative Methods}, and it was worst for \textit{Macroeconomics} and \textit{Public Economics}. 


\begin{table}[!t]
\centering
\resizebox{0.8\linewidth}{!}{
\begin{tabular}{cccc}
\toprule
                     & \textbf{Annotator 1} & \textbf{Annotator 2} & \textbf{Annotator 3} \\
\midrule
\textbf{Annotator 1} &                      & 55\%                 & 55\%                 \\
\textbf{Annotator 2} & 55\%                 &                      & 27\%                 \\
\textbf{Annotator 3} & 55\%                 & 27\%                 &    \\
\bottomrule
\end{tabular}}
\caption{Annotator Agreement Matrix Among Three Expert Annotators.}
\label{tab:iaa}
\end{table}

\begin{figure*}[!t]
    \centering  \includegraphics[width=0.7\linewidth]{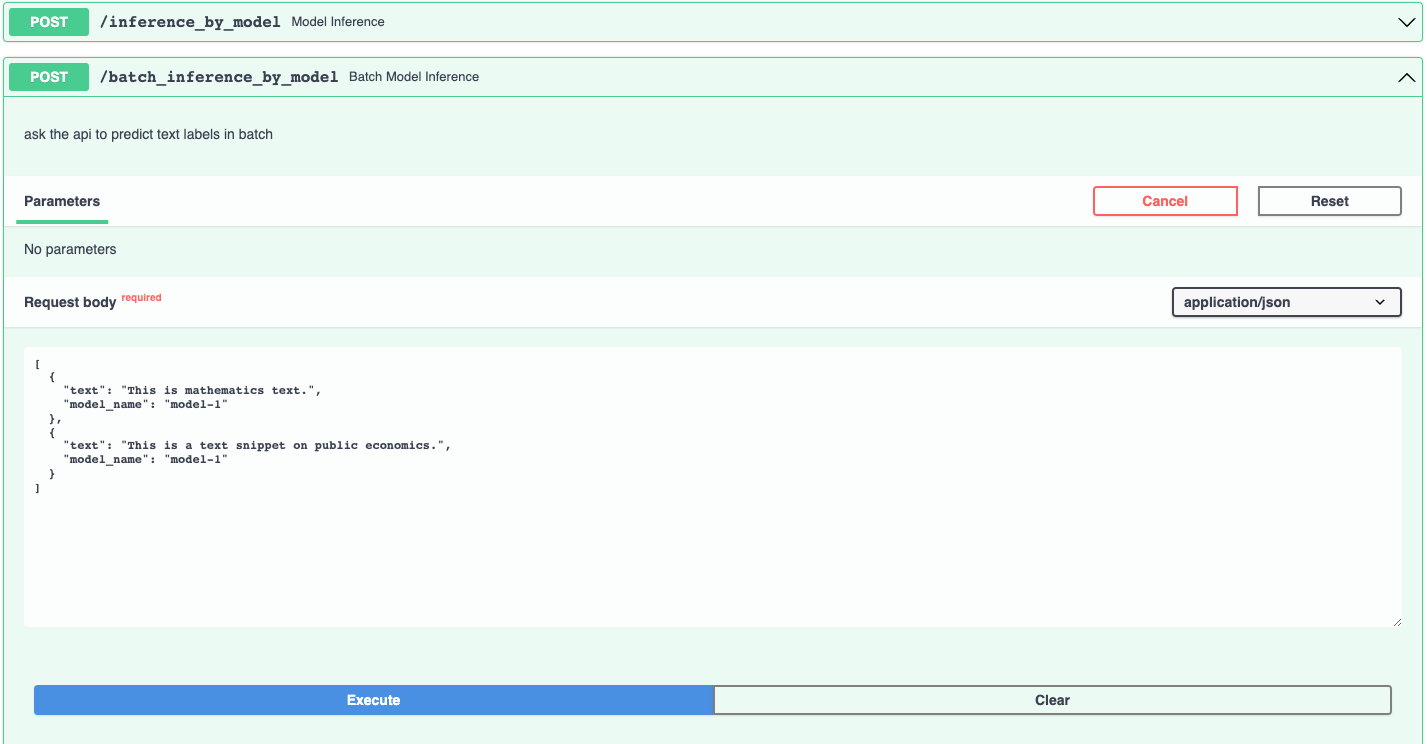}
    \caption{Inference Engine with \textit{MLflow} Integration. API: Batch Inference by Model.}
    \label{fig:inference}
\end{figure*}

\section{Inference Engine: Incorporating Annotation Results into the Existing Classification Pipeline}
\label{sec:inf}
We illustrate the pipeline using the discipline of \textit{Economics} as discussed in Section \ref{sec:implementation}. 

\subsection{Post-processing of Annotation Results}
We downloaded \href{https://gitlab.ethz.ch/raox/science-clf/-/blob/labelstudio/labelstudio/labeled_data/labelstudio_econ.json}{the annotation results in JSON} of all experts and post-processed them following the protocols below, before feeding them into the pre-trained base models of various neural networks as discussed in \citet{rao2023hierarchical}. In total, we obtained 1,068 partly overlapping annotations (incl.~"Skip", "(Dis)agree", keywords, added categories). The basic statistics on the number of instances of "Agree", "Disagree" and "Not ECON" are 498, 297, and 268, respectively.
    
The post-processing procedure is structured as follows. (1) We removed abstracts that were inadequately classified as belonging in \textit{Economics} from the sample (206 of 788 instances). Additionally, we deleted 5 instances due to bad annotations. For example, no one labeled this sample ("Skip"), or an annotator chose "Disagree" but did not choose a new category.
(2) For each remaining instance, we counted the percentages of \textit{"Agree"} and \textit{"Disagree"} verdicts relative to the label generated on the basis of MAG keywords. If strictly more experts agreed than disagreed with MAG, the original label was preserved (for 351 of the 577 valid instances).  Otherwise, we took the label suggested by the majority of annotating experts (for 226 of the 577 valid instances). 
(3) In the case of ties, we randomly picked a label from the suggested annotations (for 22 of the 226 category-renewed instances). Following this protocol, we obtained 561 instances with expert-curated labels to fine-tune the base models. 

\subsection{Fine-tuning Pre-trained Base Models}
We used the 561 labels generated by the experts as a fine-tuning set on the base models reported in \citet{rao2023hierarchical} on the discipline of \textit{Economics} (model-1). We compared the inference performances of the base model (\textbf{Model} in Table~\ref{tab:m-vs-mft}) with those of the fine-tuned model (\textbf{Model\_FT} in Table~\ref{tab:m-vs-mft}) on various neural network architectures, Deep Neural Network (DNN), Recurrent Neural Network (RNN), Convolutional Neural Network (CNN), and Transformers. To benchmark the differences in performances between \textbf{Model} and \textbf{Model\_FT}, we created a small test set from the \href{https://www.ssrn.com/index.cfm/en/}{Social Science Research Network (SSRN)}, which is a website that provides a platform for researchers to share and distribute their research papers and other scholarly work in the social sciences and other related fields. We decided to use the \textit{Economics} SSRN publications because they come with human-currated JEL categories, keywords, and abstracts.  

Concretely, we built \href{https://gitlab.ethz.ch/raox/science-clf/-/tree/labelstudio/labelstudio/ssrn_crawler}{a crawler} to download the publication space in \href{https://papers.ssrn.com/sol3/displayjel.cfm}{Economics publications in SSRN}, where all contained research articles in Economics are multi-category-indexed. This means, each publication there is indexed by at least one JEL code and it allows multiple JEL codes per publication. We could easily validate with our multi-label engine in principal, but we focus on \textit{single-label} classifications for this user study. 

To create this test set, we randomly sampled 10 instances from each of 19 JEL field classes, which resulted in a sample of 190 test instances. In the implementation of hierarchical classifications reported in \citet{rao2023hierarchical}, we have used \href{https://mlflow.org/}{MLflow} to track and manage ML experiments, with which we have saved all pre-trained base models. Now, based on them, we could seamlessly integrate model fine-tuning and inference with various models.  
The inference engine API has been implemented using \href{https://fastapi.tiangolo.com/}{FastAPI} with help from \href{https://pydantic-docs.helpmanual.io/}{Pydantic}. We illustrate the batch inference API in Figure~\ref{fig:inference}, with which users can feed the test set into various models (base or fine-tuned) and obtain predictions. In Appendix~\ref{app:inferece} we provide more details about the inference engine. 

\subsection{Benefits of Expert Annotations}
\begin{table}[!t]
\centering
\resizebox{\linewidth}{!}{
\begin{tabular}{cccccc}
\toprule
\textbf{Model} & \textbf{Model\_{FT}} & $\Delta$ (\textbf{Model\_{FT}} {-} \textbf{Model}) & \textbf{Model} = \textbf{Model\_{FT}} & \multicolumn{1}{l}{\textbf{Total}} & \textbf{Model Type}  \\
(1) & (2) & (3) & (4) & (5) & (6) \\
\midrule
54    & 58        & 4                         & 108               & \multirow{4}{*}{190}      & CNN         \\
59    & 69        & 10                        & 148               &                           & \textbf{RNN}         \\
39    & 39        & 0                         & 190               &                           & DNN         \\
31    & 33        & 2                         &  37               &                           & Transformer \\
\bottomrule
\end{tabular}
}
\caption{Results of the User Study. \textit{FT}: Fine-tuned.}
\label{tab:m-vs-mft}
\end{table}
We present the results of user studies in Table~\ref{tab:m-vs-mft}. Specifically, we inspect two types of statistics, the correct predictions of the base and fine-tuned models in Columns (1)-(2), and the identical predictions of the base and fine-tuned models in Column (4). Since each publication is multi-JEL-category-indexed, we count the prediction as "correct" if the indices include the predicted category. Column (1) is the base model trained with the model type specified in Column (6). Column (2) presents the results of the fine-tuned (supervised) model. Column (4) shows that out of a total of 190 test instances, identical predictions were generated by the base and fine-tuned models. We see that fine-tuning with user-generated results has brought benefits to all models except DNN because DNN predicts for all test examples only one class (the dominating one). RNN is the best performer when considering the benefits resulting from expert supervision, because the $\Delta$ in correct predictions has increased the most according to Column (3). Interestingly, fine-tuning a pre-trained Transformer model may not always result in a significant improvement in performance, as we see from a comparison with other base models. However, the current fine-tuning set is too small to draw firm conclusions in this regard.

\section{Discussions: Improving Annotation Efficiency}\label{sec:annotation-efficiency}
We share preliminary results of improving annotation efficiency based on the annotators' feedback.

\subsection{Similarity between Articles and Scholars}

We try to match the best suited scholars to the articles to annotate by extracting keywords from the top cited articles of these scholars and scoring them on the cosine similarity with the article keywords. The results are promising and can reduce work overhead for the scholars while improving the quality of the annotations by assigning best suited scholars for the process. More details on the implementation are given in Appendix \ref{app:scholar-sim}.

\subsection{LLM as Annotators}
In light of our commitment to total project transparency, we have opted to utilize the Vicuna 7B and 13B models \cite{vicuna}, both of which are publicly available for non-commercial use and are fine-tuned based on LLaMA \cite{llama}, explicitly tailored for QA tasks. For their predictions on single-label and multi-label classifications, see Tables~\ref{tab:vicuna_single_label} and \ref{tab:vicuna_multilabel}. The details of experimental protocols are in Appendix \ref{app:llm}. Overall, we observe that even one of the best LLMs performs poorly in the \textit{single-label} setting and has potential to be used as keyword extractor and annotate \textit{multi-label} classifications.  

\begin{table}[!t]
\centering
\resizebox{1\linewidth}{!}{
\begin{tabular}{cccccc}
\toprule
\textbf{Vicuna} & \textbf{Agree} & \textbf{Disagree} & \textbf{NOT ECON} & \textbf{Blank} & \textbf{Total} \\ 
\midrule
7B & 740 & 28 & 16 & 4 & 788 \\
13B & 122 & 648 & 7 & 11 & 788 \\
\bottomrule
\end{tabular}
}
\caption{Results of \textit{single-label} prediction. Those Vicuna outputs that do not contain the keywords "Agree", "Disagree" or "NOT ECON" are labeled as "blank".}
\label{tab:vicuna_single_label}
\end{table}

\begin{table}[!t]
\centering
\resizebox{1\linewidth}{!}{
\begin{tabular}{cccccc}
\toprule
\textbf{Vicuna} & \textbf{ECON (Y/N)} & \textbf{Cat1 (A/D/B)} & \textbf{Cat2 (A/D/B)} & \textbf{Cat3 (A/D/B)} \\ 
\midrule
7B & 417 / 297 & 157 / 146 / 114 & 122 / 199 / 96 & 5 / 18 / 10 \\
13B & 614 / 100 & 158 / 445 / 11 & 121 / 490 / 3 & 3 / 39 / 0 \\
\bottomrule
\end{tabular}
}
\caption{Results of \textit{multi-label} prediction by Vicuna. Y/N - "Yes"/"No", A/D/B - "Agree"/"Disagree"/"Blank". "Cat1,2,3" are three predicted labels by our classification system. If it predicts that an abstract does not belong to ECON, then we will no longer ask whether or not it agrees with the our model-predicted categories. Responses that do not contain the specified keywords are considered "blank". The dataset contains a total of 42 entries with non-empty Cat3. In this subset, Vicuna-13B uniformly classifies all entries to be in the ECON domain. Conversely, Vicuna-7B predicts 33 of these entries within the ECON domain.}
\label{tab:vicuna_multilabel}
\end{table}

\section{Conclusions}
In this system demonstration, we utilize a set of standard open-source software (mainly Label Studio \cite{LabelStudio}, \href{https://mlflow.org/}{MLflow} and \href{https://fastapi.tiangolo.com/}{FastAPI}) to configure an annotation and inference engine for scientific publications (SAINE). This system is built on top of hitherto largest multi-class hierarchical classification study across all disciplines in both single-label and multi-label settings (cf.~\citet{rao2023hierarchical}). We illustrate the functionality of the system with a user study in \textit{Economics} and show that the expert inputs into our system can help better understanding the classification process, which benefits the development of a stronger model in the next iteration. We plan to open-source the data and codebase and invite collaborative work in the direction of meta-science. 


\section*{Limitations}


Label Studio has some limitations in incorporating existing ML pipelines into the annotation engine, especially, when using customary code. We will discuss this with the developers at Label Studio and see how we can bring the annotation engine and the ML pipeline closer to each other. 

In terms of annotator selection, at the moment we have to select the experts for each discipline. However, we have performed experiments to rank the annotators by their field expertise and find the best annotation tasks based on the similarity between the space for academic publications and the space for articles (Appendix \ref{app:scholar-sim}). One future idea is to automatically compute an associative score between a third-party academic product such as Google Scholar and the publication space. For instance, the project \href{https://github.com/poloclub/people-map}{PeopleMap} provides interesting techniques to generate researcher profiles based on their research interests and publications taking as input the Google Scholar profile URLs of researchers. At this stage, Label Studio developers suggest that we add a self-declarative questionnaire to each annotator, which can be used as meta-data on annotators when quantifying the annotation confidence score. Due to time constraints, we have not yet added this questionnaire, as the experts in the current user study are selected by our project PI and have strong expertise in \textit{Economics}.

In terms of annotation efforts, we have benchmarked annotation quality using LLMs, which shows that human annotators are needed to control the quality. Considering our annotators' feedback that it is time-consuming to extract keywords for humans, it makes sense to use LLMs as an annotation-assisting engine for keyword extraction at this stage. We have evaluated the LLM-generated keywords: some are quite generic given the context, and others are good fits. We plan to do a systematic evaluation of LLM-generated keywords using the \href{https://paperswithcode.com/sota/document-classification-on-wos-46985}{WOS-46985 benchmark dataset}. In terms of label prediction, we see in Tables \ref{tab:vicuna_single_label} and \ref{tab:vicuna_multilabel} that Vicuna performs poorly on the \textit{single-label} task and we will need to finish a larger sample of \textit{multi-label} task to gauge its values, despite its superior performances on 100 tasks we evaluate against human performance.

\section*{Ethics Statement}
We acknowledge that our system may involve processing potentially sensitive data (such as annotator profile), and we take data privacy and ethical considerations very seriously. In accordance with ethical guidelines of "ACM Code of Ethics", we will take steps to protect the privacy of annotators once the annotation engine is in beta stage. We have also made efforts to ensure that our system and its annotations are unbiased and fair. We believe that our work will help foster greater transparency and understanding in scientific research, and we welcome collaboration and feedback from the scientific community to further advance ethical and responsible use of AI in research.

\section*{Broader Impact Statement}
Our annotation engine and inference engine can further support downstream meta-science projects. We list a few interesting questions we can answer using our pipeline (\citet{rao2023hierarchical} and the annotation and inference engine).
\begin{enumerate}[leftmargin=*, itemsep=0pt, parsep=0pt, topsep=0.5pt]
    \item \textbf{[For students.] }Which fields of research are more impactful/growing? 
    \item \textbf{[For policy makers.]} How to design education for cross-/inter-/pluridisciplinary studies? 
    \item \textbf{[For department and tenure committees.]} How to benchmark output and impact levels of an untenured scholar?
    \item \textbf{[For funding institutions.]} How to measure/quantify inter-/pluri-disciplinary standards for institutions such as \href{https://snis.ch/}{SNIS} and \href{https://www.snf.ch/en}{SNSF} which emphasize the interdisciplinarity of research?
    \item \textbf{[For librarians.]} How can one effectively organize bibliographical resources across disciplines and departments in one university?\footnote{Our team is invited to contribute to a blog post of \href{https://yearofscientometrics.ethz.ch/}{Swiss Year of Scientometrics}, a project of ETH Library, funded by \href{https://www.swissuniversities.ch/en/}{swissuniversities}.}
\end{enumerate}

We plan to add other disciplines covered by \citet{rao2023hierarchical} to our annotation engine. We would also like to incorporate subjective (self-declaration) and objective measurements (e.g., Google Scholar profile integration) into the annotation pipeline. This may help develop confidence scores of one annotation/annotator.

\section*{Acknowledgements}
We thank the colleagues at \href{https://ds3lab.inf.ethz.ch/}{DS3Lab} for providing valuable feedback when prototyping the system design. Without the strong support of our expert group at the \href{https://cae.ethz.ch/}{Chair of Applied Economics}, the user study would not have been possible. We thank Mr.~Prakhar Bhandari and Ms.~Piriyakorn Piriyatamwong for their technical support to our project. We appreciate that Label Studio has offered us an academic license for the project, which allows us to invite more experts to contribute in the long run. The user agreement and terms of an academic license are listed \href{https://labelstud.io/academic/}{here}.

\bibliography{anthology,custom}
\bibliographystyle{acl_natbib}

\appendix

\section{Our Hierarchical Classification System}
\label{app:sys}

We provide an overview of \citet{rao2023hierarchical}. The presented paper introduces a modularized three-level hierarchical classification system designed to automatically categorize scholarly publications based on their abstracts. The system operates within a hierarchical label set consisting of disciplines, fields, and subfields, enabling multi-class classification. This approach facilitates a systematic categorization of research activities, considering both knowledge production and impact through citations. The system distinguishes 44 disciplines, 718 fields, and 1,485 subfields, leveraging a vast collection of abstract snippets from the Microsoft Academic Graph. By utilizing various neural network models, such as DNNs, RNNs (using GRU), CNNs, and Transformers, through batch training, the system achieves high classification accuracy rates exceeding 90\% in both single-label and multi-label settings.

The modular design of the system allows for flexibility and easy integration of new models, with CNNs identified as the most efficient performer across the models. The system consists of three components: the first component (L1) handles discipline classification, the second component (L2) focuses on field classification, and the third component (L3) specializes in subfield classification. Each component operates based on the output of the previous level, enabling a granular categorization of research activities and capturing the interdisciplinary nature of certain topics.

In the classification process, the system assigns publications to disciplines, fields, and subfields based on their abstracts. It computes conditional probabilities to determine the relevance of each label given the previous level labels. This hierarchical approach improves the alignment of research texts with disciplines, enables automated classification, and captures interdisciplinarity.

The system incorporates both single-label and multi-label settings. In the single-label setting, each publication is assigned to a single category, while in the multi-label setting, publications can be assigned to multiple categories simultaneously. The multi-label classification assumes label independence and employs binary cross-entropy loss for training. To ensure a balanced distribution of relevant and irrelevant samples, stratified sampling is maintained for label sets.

Performance evaluation of the classification system includes metrics such as categorical accuracy, precision, and recall. The system's ability to accurately classify research texts, align them with relevant disciplines, and capture interdisciplinarity contributes to its value in indexing and analyzing scientific publications.

Overall, the proposed system, with its modular design and pretrained models, serves as a solid foundation for future applications in scientific publication indexing and analysis.

\section{Label Studio Functionalities}
\label{app:more-function}
\begin{figure*}[!ht]
    \centering
    \includegraphics[scale=0.25]{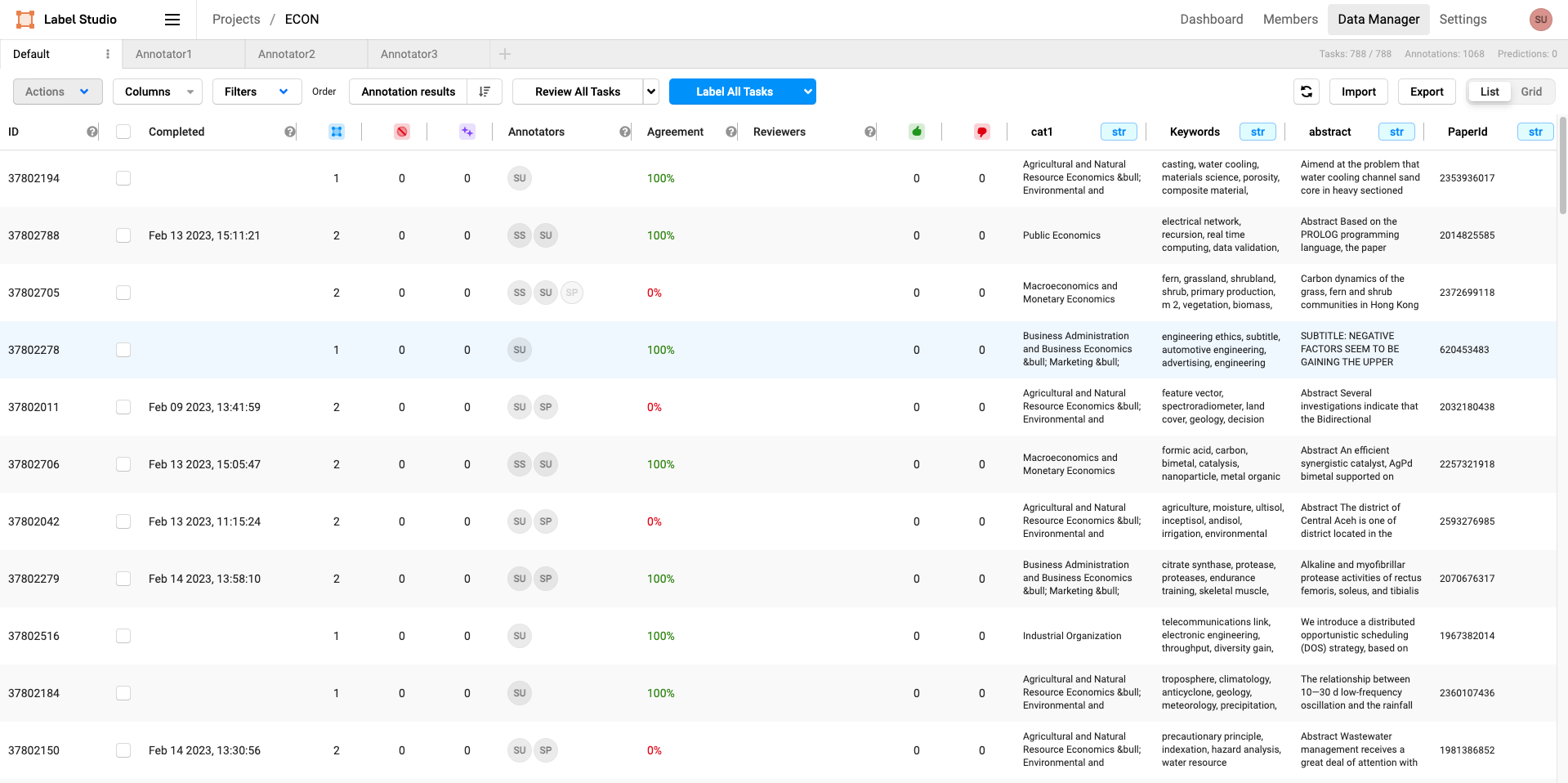}
    \caption{Administrative Panel of Annotation Tasks in Label Studio.}
    \label{fig:labelstudio_admin}
\end{figure*}

\begin{figure*}[!ht]
    \centering
    \includegraphics[scale=0.25]{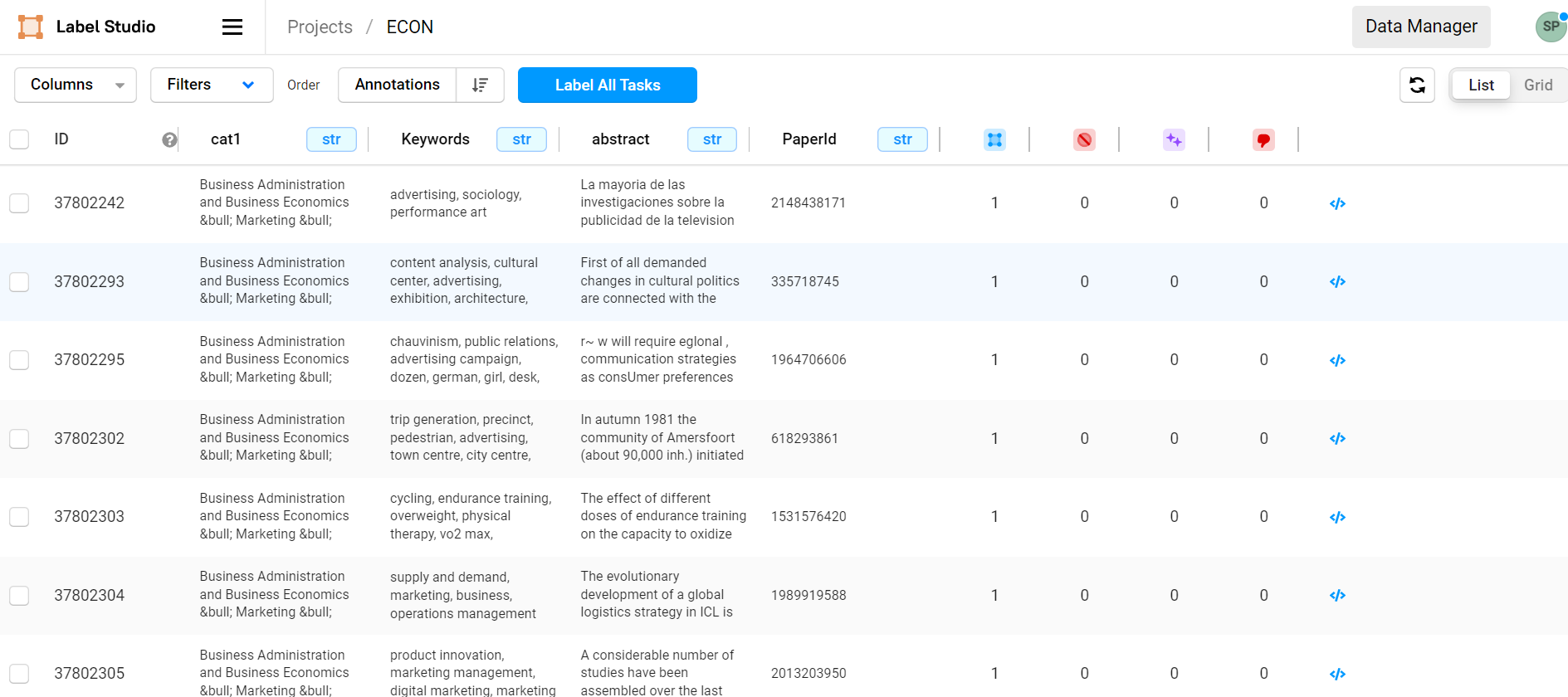}
    \caption{Annotator Panel of the Assigned Annotation Tasks in Label Studio.}
    \label{fig:labelstudio_annotator}
\end{figure*}

In Figure~\ref{fig:labelstudio_admin} we demonstrate the administrative panel of the project manager. The "Filters" and "Order (Annotation results)" tabs make it easy to inspect tasks by annotation progress (e.g., "Annotators", "Agreement", "Completed", "Total annotations per task"). 
In Figure~\ref{fig:labelstudio_annotator}, we demonstrate the user panel that we show for each expert annotator. Each annotator has no access to additional information about the annotations made by the other annotators. As an expert, one can only see how many annotations have been gathered per instance among the experts together. 

\section{Multi-label Annotation Engine}
\label{app:multi-label}
The setup of multi-label annotation is similar to the single-label setting. In the multi-label settings, annotators are required to mark "(Dis)agree" for each suggested JEL category (we provide three categories at most) and then select additional JEL categories, where multiple choices are allowed.

\begin{figure*}[!ht]
    \centering
    \begin{tabular}{c}
        \includegraphics[scale=0.25]{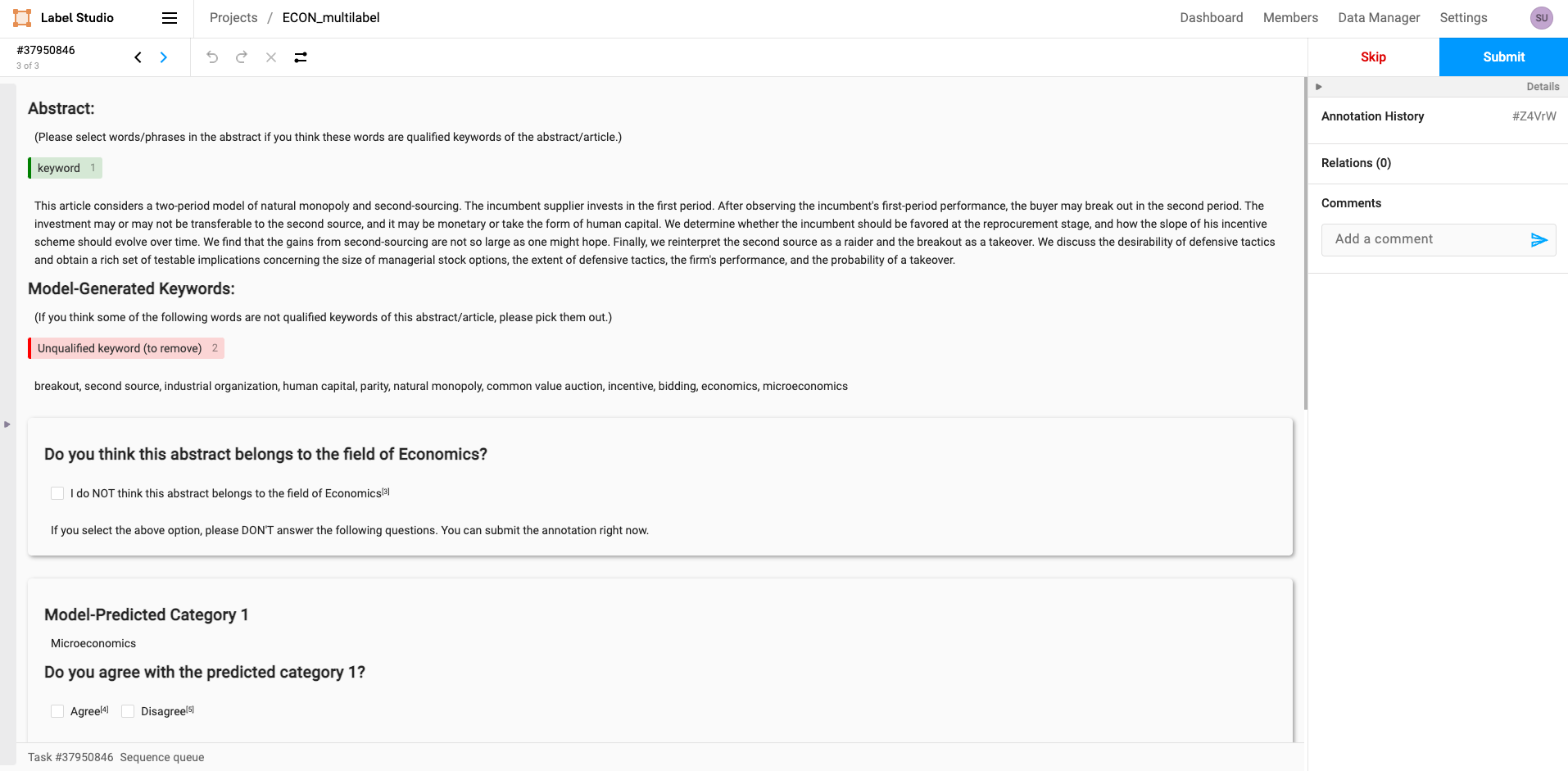} \\
        \includegraphics[scale=0.25]{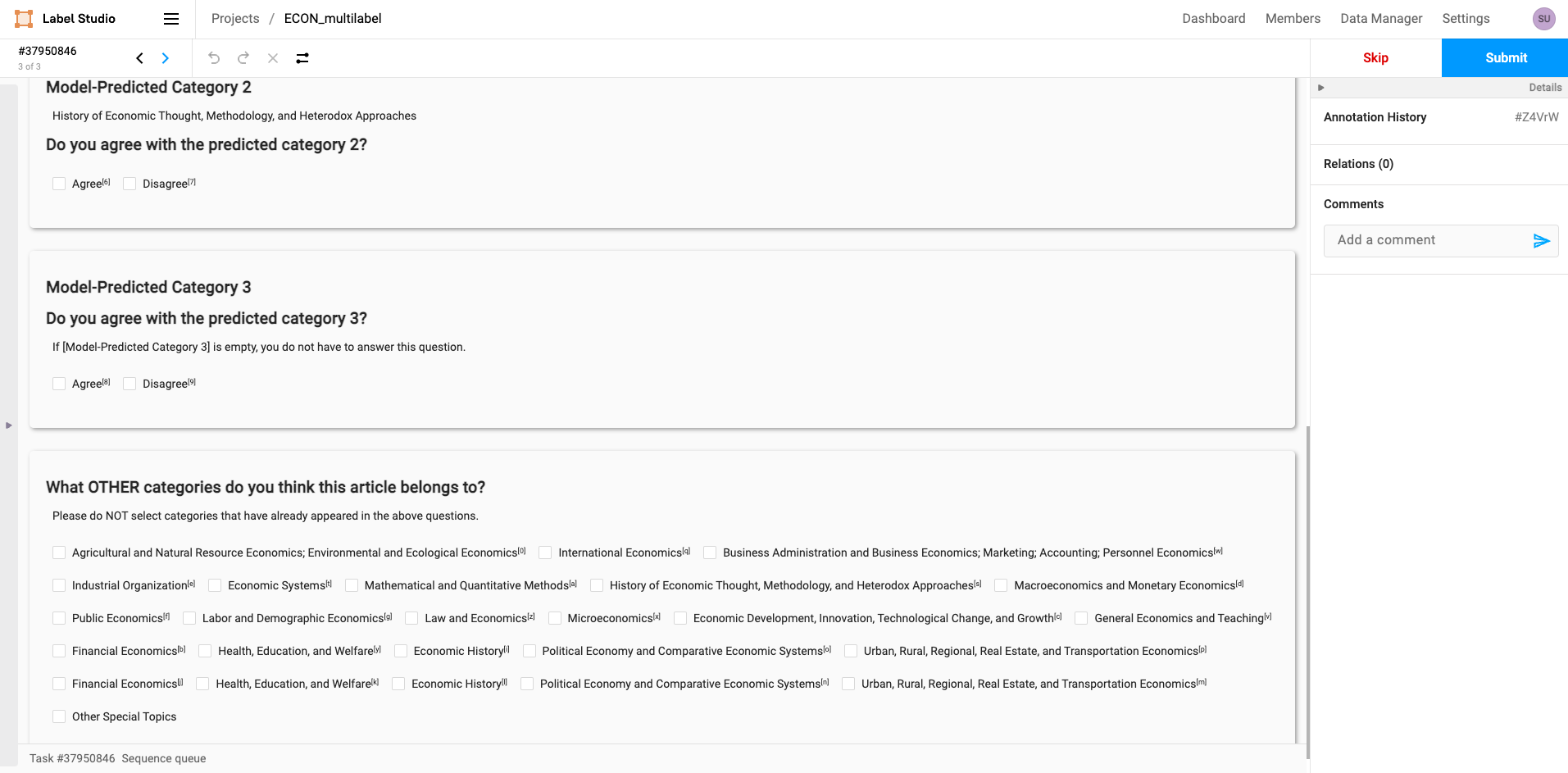} \\ 
    \end{tabular}
    \caption{Publication Annotation Engine in a Multi-label Setting.}
    \label{fig:multilabel}
\end{figure*}

\section{Inference Engine}
\label{app:inferece}
\begin{figure*}[!ht]
    \centering
    \begin{tabular}{c}
        \includegraphics[scale=0.22]{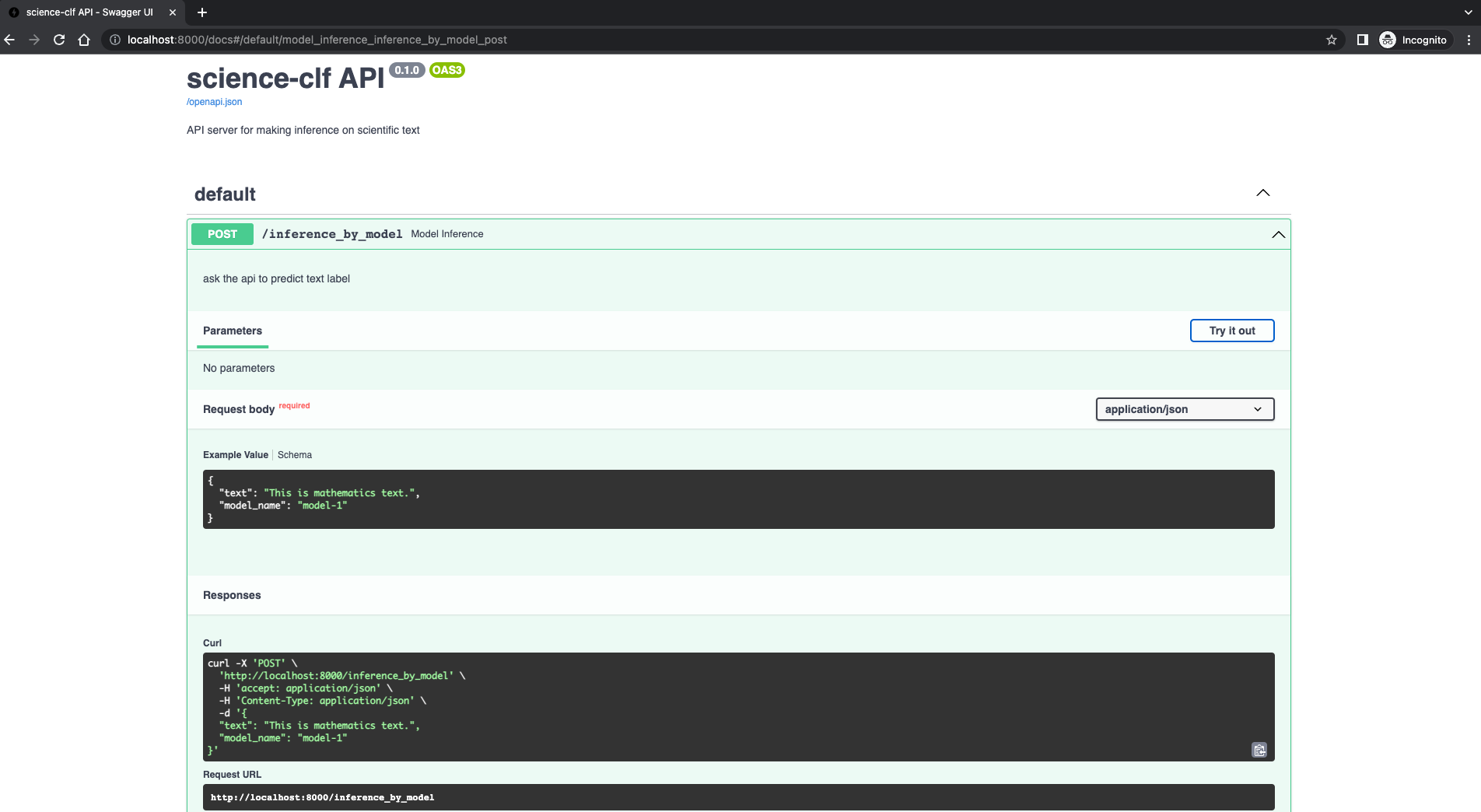} \\
        \includegraphics[scale=0.22]{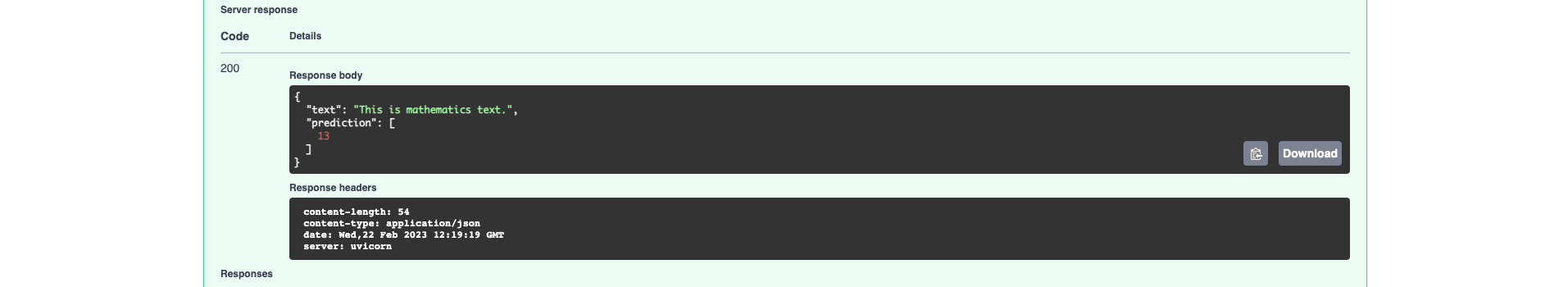} \\ 
        \textbf{(1) API: \texttt{inference\_by\_model}.} \\
        \includegraphics[scale=0.22]{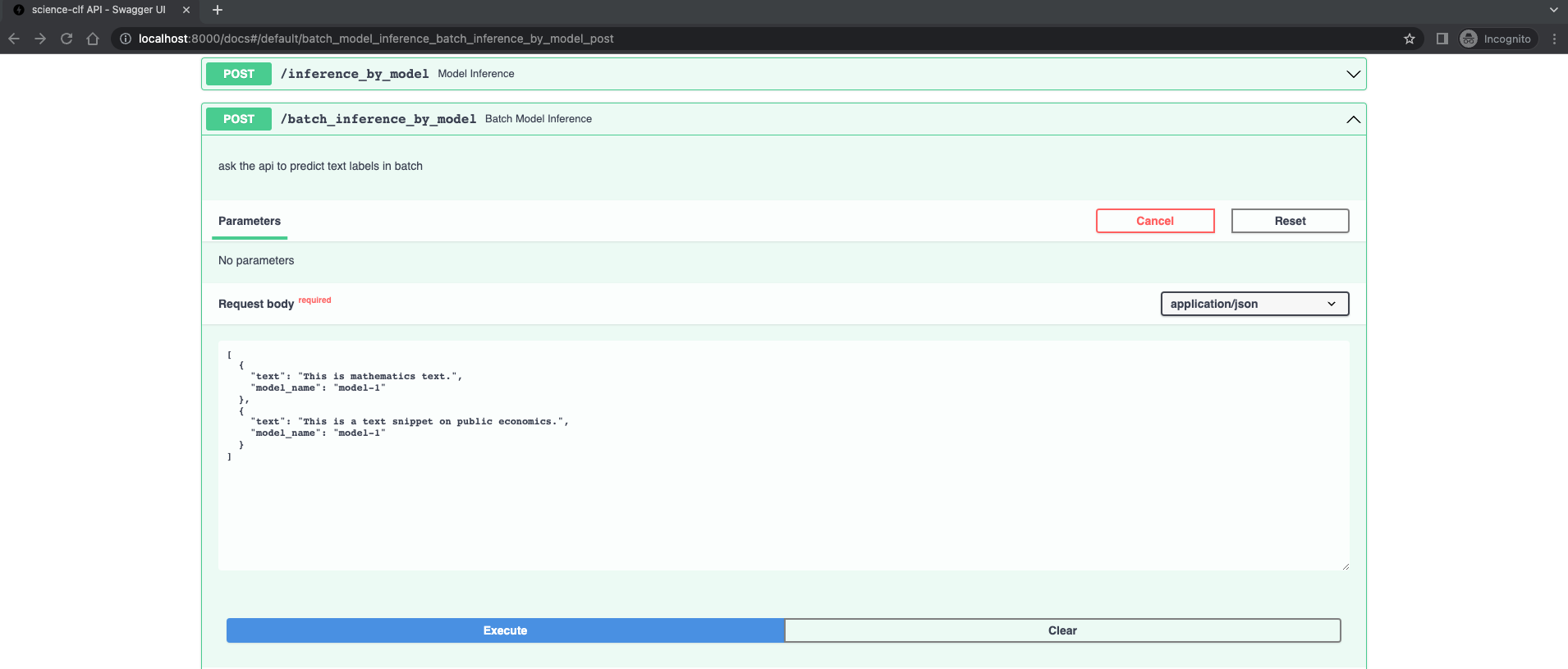} \\
        \includegraphics[scale=0.22]{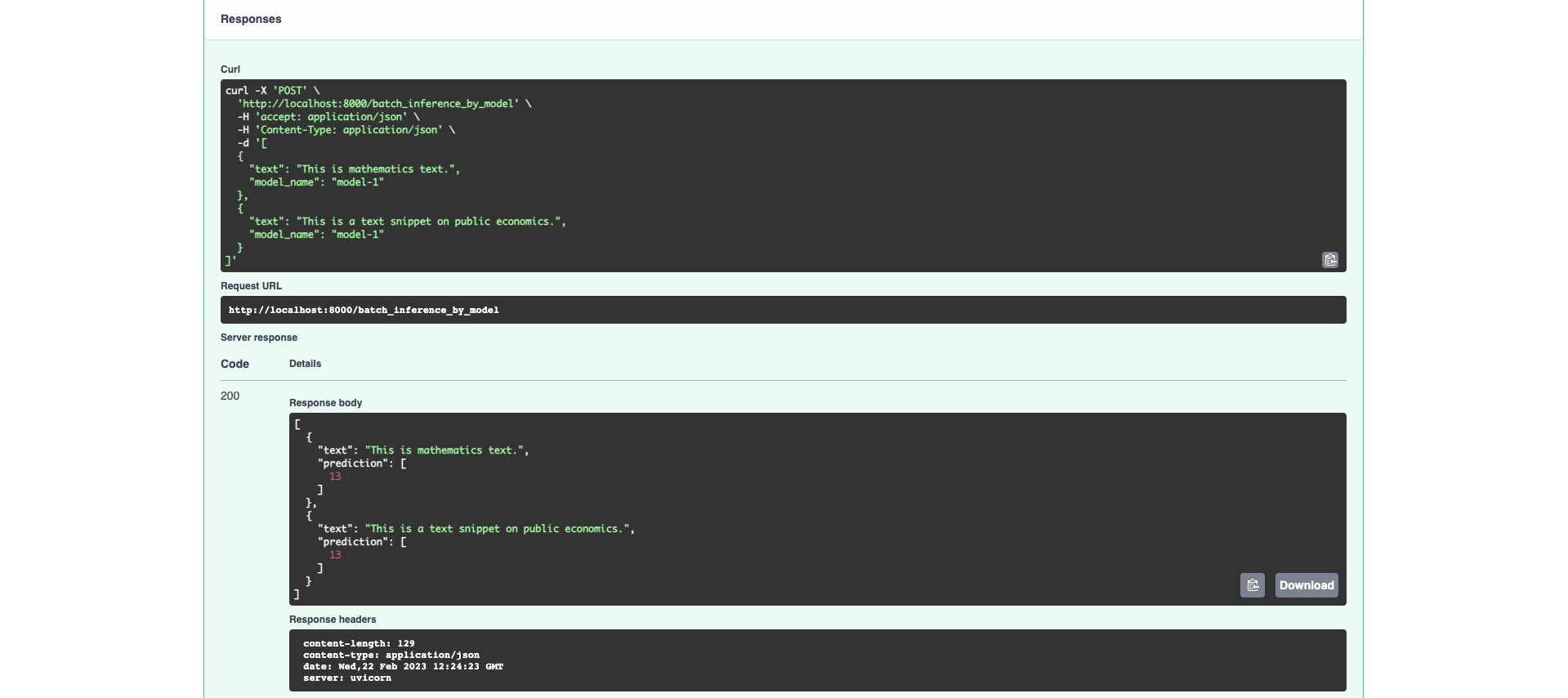} \\ 
        \textbf{(2) API: \texttt{batch\_inference\_by\_model}.} \\
    \end{tabular}
    \caption{Inference API calls.}
    \label{fig:inference-all}
\end{figure*}

We provide two types of API calls for inference, (1) \texttt{inference\_by\_model} and (2) \texttt{batch\_inference\_by\_model}. The only difference between these two API calls is that API (2) allows text-label predictions in batches, which requires a JSON sequence as input. Figure~\ref{fig:inference-all} demonstrates the user interface.

\section{Similarity between Authors and Articles}
\label{app:scholar-sim}

\subsection{Keyword Extraction}

The process of keyword extraction from the top 25 cited articles of each author and individual publication articles involves the following steps and methods. The scholars we picked are five renowned economists and use Google Scholar to download their profiles and publications. We start with the abstracts from these articles as the primary source of information.

Firstly, we perform an initial cleaning and preprocessing on these abstracts. This cleaning involves the removal of non-alphanumeric characters, conversion of text to lowercase, and tokenization of the text into individual words. We also remove common words, known as stop words, which do not contribute much to the overall meaning of the text. Finally, we discard words that are less than three characters long as these are typically not meaningful. This cleaning process results in a simplified and standardized version of the original text which is more suitable for further analysis.

Secondly, we ensure that all our data is in English to maintain consistency. For this, we employ a language detection function. If a text is not in English, we translate it using a translation pipeline, which is a model capable of accurately translating text from various languages to English. To handle potential memory issues with larger texts, we split the text into smaller chunks, translate each chunk separately, and then concatenate them back together.

The cleaned and translated text is then passed through KeyBERT \cite{grootendorst2020keybert}, a minimalistic transformer-based keyphrase extraction technique, which is used to extract keywords from the text. Apart from KeyBERT we also tried other keyword extraction techniques like YAKE \cite{campos2020yake} and RAKE \cite{csurfer2020rake-nltk}. We then compared which of these techniques extracted the best keywords by generating scores for the tasks on the benchmark data set "WOS-46985" discussed in \cite{keyword_extraction}. We found that KeyBERT got the best scores for the extracted keywords when compared to the reference data. KeyBERT uses BERT, a state-of-the-art transformer model for natural language processing, to convert words into high-dimensional vectors or embeddings. 
These embeddings capture the semantic meaning of the words and their context. KeyBERT then identifies clusters in these embeddings to find the most representative or "key" phrases.

We extract 250 keywords for each author by combining and analyzing the abstracts of their top 25 cited articles, which gives us a broad representation of their research interests. For individual publication articles, we extract 15 keywords to capture the essence of each specific article. We have tried 5, 10, 15 keywords for each article, and 15 gives the best results.

By following this methodology, we ensure the extraction of the most relevant and informative keywords for each author and individual article, providing us with a valuable understanding of the research landscape and the interests of the authors.

\subsection{Similarity Scores}

In our work, we present an innovative method that allows for a comprehensive understanding of the relationship between authors, publications, and research categories. This approach uses a function which not only identifies the top authors relevant to a particular article but also uncovers the top fields of research or "categories" connected to the article and hence its best annotator(s).

The function employs the TF-IDF (Term Frequency-Inverse Document Frequency) vectorization technique to transform text data into a numerical representation that can be processed by machine learning algorithms. For a given publication, it uses this technique to compare the article's abstract to those of top-cited authors, generating a list of the most similar authors.

Subsequently, the function identifies the top research categories linked to the publication by analyzing the keywords in its abstract. It applies the same process to the top-ranked author's 25 most cited articles. The result is a set of top categories that best align with the publication and the most relevant author, providing a deeper understanding of their research focus. This novel approach offers a multidimensional view of the research landscape, establishing clear links between authors, their publications, and research fields.

\subsection{Plotting the Author-Article Similarities}

\begin{figure*}[ht]
\centering
\includegraphics[width=1\textwidth]{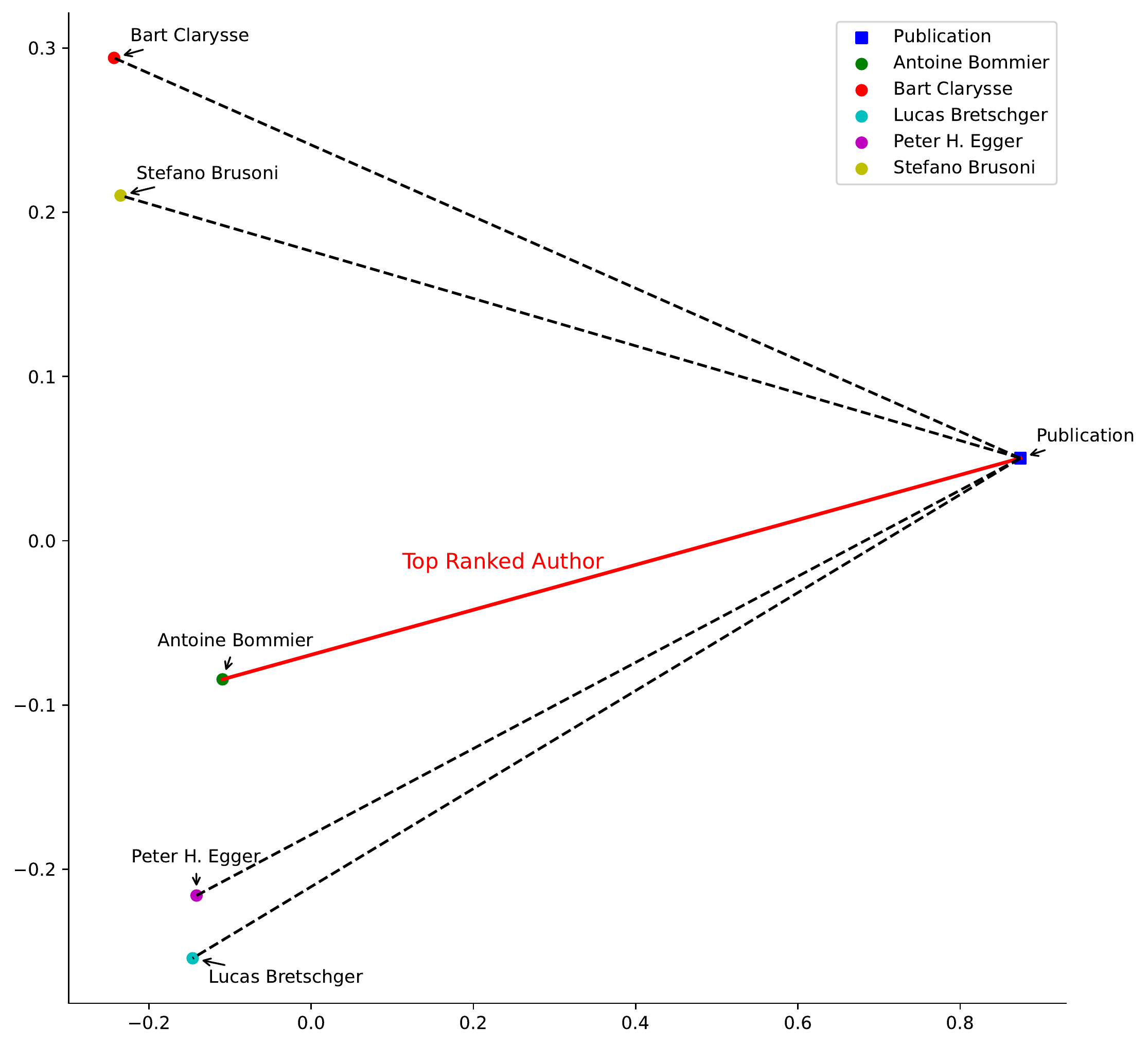}
\caption{\centering A two-dimensional representation of the semantic proximity between a specific publication and the top 25 cited author publications. Each point represents an author's work, and the spatial distribution reflects the semantic similarity to the given publication.}
\label{fig:embedding_plot}
\end{figure*}

In our research, we have developed a method for visualizing the semantic proximity between a specific publication and the top 25 cited author publications across all authors. This is accomplished through a function that maps the abstracts of the documents into a two-dimensional space using Word2Vec for word embeddings and PCA for dimensionality reduction. The resulting plot provides a graphical representation of how closely related the content of a given publication is to the influential works of various authors. In Figure \ref{fig:embedding_plot} we show an example plot produced using this method.

We invite the reader to observe the distribution of points, where the spatial proximity reflects the semantic similarity between the given publication and the authors' works. This method offers an intuitive way to understand the knowledge structure and the implicit connections between different research articles. We intend to incorporate all the above mentioned changes into Label Studio to decrease the workload of the annotators and to increase the overall efficiency and accuracy of the process. The authors will be selected through the process of ranking and choosing the one with the highest similarity score for annotating the document. The author will be given a union set of 5 categories with respect to the author's publications and the publication itself to choose from, which align the most with the publication.

\section{LLM Annotation}
\label{app:llm}
\subsection{LLM Selection}
In our work, we utilize the \href{https://github.com/lm-sys/FastChat/blob/main/docs/vicuna_weights_version.md}{v1.1 model weights} for Vicuna-7B and 13B. All inference tasks are executed on two set of RTX 3090 (24GB of memory each). The parameters employed during response generation are as follows: \texttt{max\_length} set to 100,000, \texttt{do\_sample} enabled as True, and \texttt{temperature} adjusted to 0.7. 

Our vision is to build a completely open source pipeline, so we have disregarded LLMs such as GPT-4 \cite{gpt4} which only provide API access, instead we have preferred open source alternatives such as LLaMA. We have explored non-LLaMA based LLMs such as OpenChatKit \cite{openchatkit}, but we encounter issues related to the stability of their output. We notice that these models sometimes produce inconsistent responses for the same data point (i.e., annotating one publication with keywords and labels), alternating between "Agree" and "Disagree" without providing logically coherent reasoning.

Among the multitude of LLaMA-based LLMs, we identify Vicuna as a model specifically fine-tuned for Question-Answering tasks, making it an apt choice for our project. Furthermore, Vicuna's exceptional performance, underscored by its highest Elo rating in the Chatbot Arena \cite{chatbot_arena}, convinces us to choose it as our annotator.

\subsection{Single-label}
Figure~\ref{fig:single_label_prompt} presents the standardized prompt template we employ to query the LLM  regarding its agreement with the category predicted by our model for each data point in the ECON \textit{single-label} dataset. It should be noted that certain segments of the prompts remain fixed and repetitive, a feature we refer to as "instructions". This design is necessitated by the LLM's inherent propensity to forget previous text, meaning that inputting the instruction just once may compromise the quality of responses for subsequent data points. For instance, they might cease to incorporate crucial keywords such as "Agree", "Disagree", "NOT ECON". Therefore, we find it essential to provide an instruction for each data point.

\begin{figure*}
    \centering
    \begin{tikzpicture}
  \node[rectangle, rounded corners, draw=black, fill=white, inner sep=1em] (box) {%
    \begin{minipage}{0.9\textwidth}
      \textbf{Prompt: } I have trained a machine learning llm whose input is the abstract of a scientific article, and the predicted output is its predicted category.  \\
      Candidate categories include: \texttt{\{CAT1\_NAMES\}}. \\
The abstract is: \texttt{\{abstract\}}. \\
My llm-predicted category is: \texttt{\{cat1\}}. \\
You have three options: 'Agree, Disagree, NOT ECON'. \\
Please choose ONLY ONE to output. \\
If you think this article belongs to the field of economics and the category predicted by the llm is correct, please output 'Agree' and give reasons; \\
Otherwise, If you think this article belongs to the field of economics but the category predicted by the llm is incorrect, please output 'Disagree', state which category it should belong to according to your opinion, and then provide reasons; \\
Otherwise, If you think this article does not belong to the field of economics, output 'NOT ECON', and give reasons. \\
\\
      \textbf{Response: }  Disagree, Business Administration and Business Economics; Marketing; Accounting; Personnel Economics. \\
While the article does discuss adsorption/desorption behavior in alluvial soils, it also involves other economic factors such as cation exchange capacity and clay content, as well as the amount of Se desorbed in different forms. These factors suggest that the article may be more relevant to the field of Business Administration and Business Economics, which encompasses topics such as marketing, accounting, and personnel economics. Additionally, the article does not seem to fit squarely into the category of Agricultural and Natural Resource Economics \& Environmental and Ecological Economics, as it is not solely focused on environmental or natural resource issues.
    \end{minipage}
  };
\end{tikzpicture}
    \caption{LLM annotation template for single-label data. \texttt{\{CAT1\_NAMES\}} should be replaced with a string list \texttt{[Agricultural and Natural Resource Economics; Environmental and Ecological Economics, Industrial Organization, ...]}, which contains 18 elements (i.e., candidate categories). Taking Paper ID 2279482037 from the ECON single label dataset as an example, replace \texttt{\{abstract\}} with \textit{Laboratory studies were conducted to investigate the adsorption/desorption behaviour ...} and replace \texttt{\{cat1\}} with \textit{Agricultural and Natural Resource Economics; Environmental and Ecological Economics}. Response is the output of Vicuna-13B for this data point.}
    \label{fig:single_label_prompt}
\end{figure*}

\subsection{Multi-label}
For the ECON \textit{multi-label} dataset, we engage the LLM with up to five prompts for each data point (Figure~\ref{fig:multilabel_prompt}). The first prompt asks whether the given abstract is relevant to the field of Economics. If the answer is negative, we terminate further inquiry. However, if the LLM confirms the economic relevance, we proceed to query the model's agreement with up to three categories our model had previously predicted. Lastly, we ask the LLM about any additional categories to which it believes the abstract may belong, beyond those predicted by our model. This final query is intended primarily as a preparatory measure for future keyword analysis. Similarly, the persistent recurrence of identical instructions within the prompt template is designed to mitigate the issue of forgetfulness inherent in the LLM. When we compare the output of the LLM with those of the human annotators on 100 annotations, we see a significant overlap in the categories allotted using both methodologies, which shows the potential to employ LLMs for multi-label tasks. 

\begin{figure*}
    \centering
    \begin{tikzpicture}
  \node[rectangle, rounded corners, draw=black, fill=white, inner sep=1em] (box) {%
    \begin{minipage}{0.9\textwidth}
      \textbf{1st Prompt: } I have trained a machine learning llm whose input is the abstract of a scientific article, and the predicted output is its predicted categories (up to 3). \\
Candidate categories include: \texttt{\{CAT1\_NAMES\}}. \\
The abstract is: \texttt{\{abstract\}}". \\
Do you think this abstract belongs to the field of Economics? \\
If so, output 'Yes'; if not, output 'No'. \\
      \textbf{1st Response: } Yes. \\

      \textbf{2nd Prompt: } The abstract is: \texttt{\{abstract\}} \\
      Do you think this abstract belongs to the first model-predicted category \texttt{\{cat1\}}? If you agree with the first model-predicted category, please output 'Agree' and the reason; if you do not agree with the first model-predicted category, please output 'Disagree' and the reason.\\
      \textbf{2nd Response: }  Agree. The abstract describes a theoretical model of a monopoly firm that is introducing a new product with an unknown quality level to consumers. The model predicts that the firm will use price as a signal of quality, and that the price will monotonically increase with the quality level. This model falls within the category of microeconomics, which is concerned with the behavior of firms and consumers in markets, and how government policies can influence that behavior. \\

      \textbf{3rd Prompt: } The abstract is: \texttt{\{abstract\}}  \\
      Do you think this abstract belongs to the second model-predicted category \texttt{\{cat2\}}? If you agree with the second model-predicted category, please output 'Agree' and the reason; if you do not agree with the second model-predicted category, please output 'Disagree' and the reason. \\
      \textbf{3rd Response: }   Disagree. The abstract does not appear to belong to the category of "History of Economic Thought, Methodology, and Heterodox Approaches." It is more closely related to the category of "Consumer Behavior and Demand Analysis" or "Market Structure and Pricing." \\

      \textbf{4th Prompt}: Are there any other categories that you think are more suitable for this abstract, besides \texttt{\{cat1\} \{cat2\} \{cat3\}}? \\
If so, please output some other categories among candidate categories \texttt{\{CAT1\_NAMES\}} and the reasons. If not, please output the reason why not. \\
    \textbf{4th Response}:   Some other categories that might be suitable for this abstract could include: \\
* Agricultural and Natural Resource Economics: This category could be a good fit because the abstract discusses the impact of technology on the agricultural sector, including the use of drones for crop monitoring and automated harvesting. \\
* Environmental and Ecological Economics: This category could also be a good fit because the abstract mentions the potential for drone technology to improve crop yield and reduce the use of pesticides, which could have positive environmental impacts. \\
However, it is also possible that the abstract could fit into multiple categories, depending on the specific focus of the research. For example, the abstract could be categorized as both Agricultural and Natural Resource Economics and Environmental and Ecological Economics. \\
Overall, the categories of Business Administration and Business Economics, Mathematical and Quantitative Methods, Public Economics, and International Economics may not be as directly relevant to the focus of the research, although there could still be some overlap depending on the specific focus of the study. 
      
    \end{minipage}
  };
\end{tikzpicture}
    \caption{LLM annotation template for multilabel data. \texttt{\{CAT1\_NAMES\}} should be replaced with the same string list described in Figure~\ref{fig:single_label_prompt}. Taking Paper ID 1497381249 from the ECON multilabel dataset as an example, replace \texttt{\{abstract\}} with \textit{This paper analyzes the role of price as a signal of the quality of a monopoly firm's new product ...}, \texttt{\{cat1\}} with \textit{Microeconomics}, and \texttt{\{cat2\}} with \textit{History of Economic Thought, Methodology, and Heterodox Approaches}. Since our model only provides two categories, \texttt{\{cat3\}} is an empty string and no prompt is given. All responses are the outputs of Vicuna-13B for this data point.}
    \label{fig:multilabel_prompt}
\end{figure*}

\end{document}